\documentclass[12pt,a4paper]{article}

\newtheorem{lemma}{Lemma}{\bf }{\it }
\newtheorem{theorem}{Theorem}{\bf }{\it }
\newcommand{\np}{\mbox{NP}}

\title{Designing SAT for HCP}
\author{Anatoly D. Plotnikov\thanks{Vinnitsa Institute of Regional
Economics and Management, e-mail: aplot@tom.vinnica.ua}}
\date{}

\begin{document}
\maketitle

\begin{abstract}
For arbitrary undirected graph $G$, we are designing SATISFIABILITY problem 
(SAT) for HCP, using tools of Boolean algebra only. The obtained SAT be the 
logic formulation of conditions for Hamiltonian cycle existence, and use $m$ 
Boolean variables, where $m$ is the number of graph edges. This Boolean 
expression is true if and only if an initial graph is Hamiltonian. 
That is, each satisfying assignment of the Boolean variables determines a 
Hamiltonian cycle of $G$, and each Hamiltonian cycle of $G$ corresponds to 
a satisfying assignment of the Boolean variables. In common case, the obtained 
Boolean expression may has an exponential length (the number of Boolean 
literals).
\end{abstract}

\section{Introduction}

A SATISFIABILITY problem (SAT) be historically the first \np-complete problem. 

Classically, SAT is formulated the following way.

Let the $y_{i}$ ($i=\overline {1, m}$) be some propositions, and 
$f(y_{1}, y_{2}, \ldots , y_{m})$ be a compound proposition constructed from
the $y_{i}$'s. Further we assume that the value 0 for a proposition $y_{i}$ 
means that the proposition is false, and 1 means that the proposition 
$y_{i}$ is true.

Let there be a set $B_{m}=\{0, 1\}^{m}$, where 0 and 1 mean false and true
respectively. A mapping $f$: $B_{m}\rightarrow \{0, 1\}$ is called a
{\it Boolean function} on $m$ variables.

An element $\sigma=(\sigma_{1}, \sigma_{2}, \ldots , \sigma_{m})\in B_{m}$, 
where $\sigma_{i}\in \{0,1\}$ ($i=\overline{1, m}$), we shall call an 
{\it assignment} of variables for the function $f(y_{1}, y_2, \ldots , 
y_{m})$. If $f(\sigma)=1$ then the element $\sigma\in B_{m}$ we shall be call 
a {\it satisfying} assignment of variables for the Boolean function 
$f(y_{1}, y_{2}, \ldots , y_{m})$.

We denote $y_{i}^{\sigma_{i}}$ = $y_{i}$ if $\sigma_{i}=1$, and 
$y_{i}^{\sigma_{i}}$ = $\bar y_{i}$ if $\sigma_{i}=0$ ($i=\overline {1, m}$).
An element $y_{i}^{\sigma_{i}}$ is called a {\it literal}. The literals 
$y^{0}$ and $y^{1}$ we call {\it contrary}.

Any conjunction of $r$ ($r\leq m$) different non-contrary literals 
\begin{displaymath}
K=y_{i_{1}}^{\sigma_{i_{1}}}\wedge y_{i_{2}}^{\sigma_{i_{2}}}\wedge \cdots 
\wedge 
y_{i_{r}}^{\sigma_{i_{r}}},
\end{displaymath}
is called {\it elementary}. The elementary conjunction is equal to 1
if all its components are equal to 1. 

Let $K_{1}, K_{2}, \ldots , K_{s}$ be elementary conjunctions. Then
a disjunction
\begin{equation}
\label{dnf}
f=K_{1}\vee K_{2}\vee \cdots \vee K_{s} 
\end{equation}
is called the {\it disjunction normal form (DNF)}. Obviously, that DNF 
(\ref{dnf}) is equal to 1 if at least one of its components is equal to 1.

Any disjunction different non-contrary literals
\begin{displaymath}
D=y_{i_{1}}^{\sigma_{i_{1}}}\vee y_{i_{2}}^{\sigma_{i_{2}}}\vee \cdots \vee 
y_{i_{p}}^{\sigma_{i_{p}}},
\end{displaymath}
is also {\it elementary}. The elementary disjunction is equal to 1 is at 
least one of literals is equal to 1.

Let $D_{1}, D_{2}, \ldots , D_{h}$ be the elementary disjunctions. Then
a conjunction
\begin{equation}
\label{knf}
f=D_{1}\wedge D_{2}\wedge \cdots \wedge D_{h} 
\end{equation}
is called the {\it conjunction normal form (CNF)}. Obviously, that CNF 
(\ref{knf}) is equal to 1 if each of component disjunction is equal to 1.

Let there be some Boolean function in the form (\ref{dnf}) or (\ref{knf}). 
It is required to find values variables for (\ref{dnf}) or (\ref{knf})
such for which $f$ is true.

In common case it is optional that the Boolean function $f$ was represented
as the conjunction or disjunction normal form. Any other form of 
representation of the Boolean function is possible.

SAT may be considered as the logic model of any problem of the class
\np. It does not need to consider similar model simply as a rule of producing
an instance of SAT from an instance of some \np-problem. Obviously, that SAT 
be the logic formulation of conditions for existence of the HCP solution.

In this paper, we design the logic expression for existence of a Hamiltonian
cycle in an arbitrary undirected graph.

\section{The logic expression for HCP}

Consider a class $L_{n}$ of undirected graphs without loops and multiple
edges with $n$ vertices.

Let $G = (X, E)\in L_{n}$, where $X = \{x_{1},\ldots, x_{n}\}$ is the set 
of graph vertices and $E$ be the set of unordered pairs of $X$, called 
{\it edges}. 

A {\it Hamiltonian cycle} in a graph $G$ be a cycle, visiting each graph
vertex exactly once\footnote{Indefinable concepts see in \cite{west}.}. 
A graph $G\in L_{n}$ is Hamiltonian if it has a Hamiltonian cycle. HCP is 
\np-complete problem (see, for example, \cite{compendium, gary-johnson}).

Construct SAT for HCP.

Previously, we make some remarks.

As it is already mentioned above, SAT may be considered as a logical model 
of any of \np-problems. Therefore, we proceed from several assumptions 
as it is done in the time of construction of any mathematical model.

For example, let there be a set family $S=\{S_{1}$, \ldots, $S_{m}\}$. It is
required to find a transversal of $S$. In this case we mean that each of sets 
$S_{i}\in S$ ($i=\overline {1, m}$) is not empty.

Similar, we shall proceed from several ``natural'' assumptions when we
shall design SAT for HCP. Obviously, if some of them are not fulfilled
then SAT has no satisfying assignments, and the corresponding graph is not
Hamiltonian.

SAT for HCP we construct as a conjunction of two Boolean expressions:
\begin{equation}
\label{all}
F=F_{1}\wedge F_{2}.
\end{equation}

To formulate each of the Boolean expressions $F_{1}$ and $F_{2}$ we shall 
introduce the Boolean variables.

A Hamiltonian cycle $C_{H}$ of $G$, when it exists, it is worth to represent
by a {\it totality} of $n$ edges:
\begin{displaymath}
C_{H}=\{e_{i_{1}}, e_{i_{2}}, \ldots, e_{i_{n}}\}.
\end{displaymath}

Consider an arbitrary vertex $x\in X$ of $G$, having a local degree $deg(x)$. 
Let the edges $e_{i_{1}}, e_{i_{2}}, \ldots , e_{i_{deg(x)}}$ of $G$ be 
incident the vertex $x$.

\vspace{1pc}

{\bf The first assumption, which we proceed from, consists that 
$deg(x)\geq 2$ for all $x\in X$.}

\vspace{1pc}

The made assumption is evident since a graph $G$ is not two-connected, and has
no a Hamiltonian cycle if there are less than two vertices that is incident 
some vertex of $G$.

An unique Boolean variable $y_{i_{q}}$ is assigned to each edge $e_{i_{q}}$ 
$(q=\overline {1, deg(x)})$. We shall suppose that $y_{i_{q}}=1$ if 
$e_{i_{q}}\in C_{H}$, and $y_{i_{q}}=0$ otherwise.

Let the edges $e_{i_{1}}$, $e_{i_{2}}$ be incident to the vertex $x\in X$,
and belong to a Hamiltonian cycle $C_{H}$ of $G$. Obviously, then a 
conjunction
\begin{equation}
\label{cat}
K=y_{i_{1}}\wedge y_{i_{2}}\wedge {\bar y}_{i_{3}}\wedge \cdots \wedge {\bar 
y}_{i_{deg(x)}},
\end{equation}
equal to 1.

In common case, for a vertex $x\in X$ we can compose
\begin{displaymath}
t=\left( \begin{array}{c} deg(x)\\ 2\\ \end{array} \right) = 
{{deg(x)\cdot (deg(x)-1)}\over 2}
\end{displaymath}
conjunctions in the form (\ref{cat}), each of which contains two Boolean
variable without negation exactly. Let $K(x)$ = $\{K_{1}, K_{2}, \ldots , 
K_{t}\}$ be a set of all similar conjunctions.

We assign a disjunction
\begin{displaymath}
d(x)=\bigvee_{\forall K_{g}\in K(x)} K_{g}
\end{displaymath}
of conjunctions in the form (\ref{cat}) to each vertex $x\in X$ of $G$.

Thus, for any graph $G\in L_{n}$ we may determine a Boolean expression
\begin{equation}
\label{fun}
F_{1}=d(x_{1})\wedge d(x_{2})\wedge \cdots \wedge d(x_{n}).
\end{equation}

Let there is a set $W$ of cycles $C(X_{1})$, \ldots , $C(X_{w})$ of $G=(X, 
E)$. The set $W$ is called a {\it partition} of $G$ into disjoint cycles if 
$X_{i}\not= \oslash$ for all $i=\overline {1, w}$,
$X_{i}\cap X_{j}=\oslash$ ($i\not=j$) for all $i, j\in \{1, \ldots , w\}$ and 
$\bigcup_{i=1}^{w} X_{i}=X$. Else, in this case the set $W$ is called a 
{\it 2-factor} of $G$ \cite{harary, west}, or a {\it vertex disjoint cycle 
cover} \cite{compendium}.

\begin{lemma}
\label{fn1}
The Boolean expression (\ref{fun}) is true if and only if a graph vertices are
splited into disjoint cycles. 
\end{lemma}

\noindent
{\bf Proof}.
Obviously, if a graph vertices are splited into disjoint cycles then the 
expression (\ref{fun}) is true.

On the second hand, a satisfying assignment of Boolean variables (\ref{fun}) 
determines some totality $E_{1}$ of graph edges. By construction of 
(\ref{fun}), for any vertex of $G$ in $E_{1}$ there exists two edges exactly 
which are incident to the given vertex. Therefore, the totality $E_{1}$ 
determines some partition $W$ of $G$ into disjoint cycles.$\circ$

\vspace{1pc}

Note that the expression (\ref{fun}) may be true if $G$ is unconnected,
for instance, if it consists of two unconnected distinct cycles.

The question raises: how we can take into account two-connected of $G$ in
SAT for HCP?

Consider some cycle $C(S)$ of $G$, where $S$ is the vertex set of the cycle.
The edge set of $G$, for which one and only one terminal vertex is incident 
to some vertex of $S\subset X$, we denote by $E(S)$. Further, let $R(S)$ be 
the set of edge pairs of $E(S)$ such that they have no common vertex in $S$.

\vspace{1pc}

{\bf The second of our assumptions consists that for any cycle $C(S)$ of $G$, 
the set $R(S)$ is not empty if $S\not=X$.} 

\vspace{1pc}

That is, if some cycle of $G$ does not contain all graph vertices then there 
exists at least two edges for its, each of which has one terminal vertex 
only that is incident to distinct vertex of this cycle.

Else speaking, if the graph is Hamiltonian then any Hamiltinian cycle must
goes into any cycle $C(S)$ ($S\subset X$), and goes out from the cycle.

As the first assumption, the second assumption is also natural since
if it is fulfilled then the graph, obviously, is not two-connected, and,
hence, it has no a Hamiltonian cycle.

Let $C(S)$ be some cycle of a Hamiltonian graph $G$.

We shall assign a conjunction $y_{i_{1}}\wedge y_{i_{2}}$ to each edge pair 
$(e_{i_{1}}$, $e_{i_{2}})\in R(S)$. Clearly that this conjunction 
be absent in $F_{1}$ if the corresponding edges have no common vertices.

For the cycle $C(S)$ we compose a disjunction
\begin{displaymath}
D(S)=\bigvee (y_{i_{1}}\wedge y_{i_{2}})
\end{displaymath}
of conjunctions, each of which corresponds to one of elements of the set
$R(S)$.

Then a set of disjunctions $D(S)$ for all cycles $C(S)$ of $G$ induce 
an expression
\begin{equation}
\label{cc}
F_{2}=\bigwedge D(S).
\end{equation}

\begin{theorem}
SAT, constructed by expression \ref{all}, has an exponential number of
conjunctions.
\end{theorem}

\noindent
{\bf Proof.} An exponential number of conjunctions in the expression
for SAT follows from an exponential number cycles of $G$ (see, for example, 
\cite{reingold}).

\begin{theorem}
The expression $F=F_{1}\wedge F_{2}$ is true if and only if a graph $G$ 
is Hamiltonian.
\end{theorem}

\noindent
{\bf Proof.} Indeed, let $G\in L_{n}$ be a Hamiltonian graph. By Lemma 
\ref{fn1}, the expression $F_{1}$ is true. Besides, if $C(S)$ be a cycle
such that $S\subset X$, $S\not= X$ then at least two edges of $E(S)$ belong
to a Hamiltonian cycle, that is, the expression $F_{2}$ is also true.

Converse, if the expression $F=F_{1}\wedge F_{2}$ is true then we have a
partition of $G$ into disjoint cycles. If we suppose that this partition
contains more than one cycle then we have contradiction since the
expression $F_{2}$ is true.$\circ$

\vspace{1pc}

\noindent
{\bf Example.}

Let there be a graph, shown on Fig. \ref{ex} (a). Construct SAT for
HCP of the given graph.

\begin{figure}
\centering
\unitlength 1.00mm
\linethickness{0.1pt}
\begin{picture}(68.89,34.22)
\put(20.00,10.00){\line(-1,1){10.00}}
\put(10.00,20.00){\line(1,1){10.00}}
\put(20.00,30.00){\line(1,-1){10.00}}
\put(30.00,20.00){\line(-1,-1){10.00}}
\put(10.00,20.00){\line(1,0){20.00}}
\put(20.00,20.00){\line(0,1){10.00}}
\put(20.00,10.00){\circle*{2.50}}
\put(10.00,20.00){\circle*{2.50}}
\put(20.00,20.00){\circle*{2.50}}
\put(30.22,20.00){\circle*{2.50}}
\put(20.00,30.00){\circle*{2.50}}
\put(20.00,6.00){\makebox(0,0)[cc]{1}}
\put(6.00,20.00){\makebox(0,0)[cc]{2}}
\put(20.00,34.22){\makebox(0,0)[cc]{3}}
\put(34.00,20.00){\makebox(0,0)[cc]{4}}
\put(20.00,16.44){\makebox(0,0)[cc]{5}}
\put(13.33,26.50){\makebox(0,0)[cc]{$a$}}
\put(21.33,24.89){\makebox(0,0)[cc]{$b$}}
\put(26.00,26.50){\makebox(0,0)[cc]{$c$}}
\put(15.33,21.80){\makebox(0,0)[cc]{$d$}}
\put(24.44,21.80){\makebox(0,0)[cc]{$e$}}
\put(12.67,13.20){\makebox(0,0)[cc]{$f$}}
\put(26.44,13.20){\makebox(0,0)[cc]{$g$}}
\put(55.00,10.00){\line(-1,1){10.00}}
\put(45.00,20.00){\line(1,1){10.00}}
\put(55.00,30.00){\line(1,-1){10.00}}
\put(65.00,20.00){\line(-1,-1){10.00}}
\put(45.00,20.00){\line(1,0){20.00}}
\put(55.00,10.00){\circle*{2.50}}
\put(45.00,20.00){\circle*{2.50}}
\put(55.00,20.00){\circle*{2.50}}
\put(65.00,20.00){\circle*{2.50}}
\put(55.00,30.00){\circle*{2.50}}
\put(55.00,6.00){\makebox(0,0)[cc]{1}}
\put(40.00,20.00){\makebox(0,0)[cc]{2}}
\put(55.00,35.00){\makebox(0,0)[cc]{3}}
\put(68.89,20.00){\makebox(0,0)[cc]{4}}
\put(54.89,16.44){\makebox(0,0)[cc]{5}}
\put(48.22,26.50){\makebox(0,0)[cc]{$a$}}
\put(60.89,26.50){\makebox(0,0)[cc]{$c$}}
\put(50.22,21.80){\makebox(0,0)[cc]{$d$}}
\put(59.33,21.80){\makebox(0,0)[cc]{$e$}}
\put(47.56,13.20){\makebox(0,0)[cc]{$f$}}
\put(61.33,13.20){\makebox(0,0)[cc]{$g$}}
\put(20.00,1.00){\makebox(0,0)[cc]{(a)}}
\put(55.00,1.00){\makebox(0,0)[cc]{(b)}}
\end{picture}
\caption{}
\label{ex}
\end{figure}

The edges of the given graph are assigned to the Boolean variables $a$, $b$, 
\ldots , $g$. It is not difficult to see that the expression $F_{1}$ has the
following form (further we cast out the symbol of the conjunction in the 
expressions):

\begin{displaymath}
F_{1}=(fg)(ad{\bar f}\vee a{\bar d}f\vee {\bar a}df))(ab{\bar c}\vee 
a{\bar b}c\vee {\bar a}bc)(ce{\bar g}\vee c{\bar e}g\vee {\bar c}eg)
(bc{\bar d}\vee b{\bar c}d\vee {\bar b}cd)
\end{displaymath}

The given graph has the following cycles, each of which does not contain 
all vertices of the graph:
1--2--5--4, 2--3--5, 2--3--4--5, 3--4--5.

The disjunction for the cycle 1--2--5--4 has a form: $(ab\vee ac\vee bc)$,
for the cycle 2--3--5: $(ce\vee cf\vee ef)$, for the cycle 2--3--4--5: $(fg)$,
and, at last, for the cycle 3--4--5: $(ad\vee ag\vee dg)$.

Thus, the expression $F_{2}$ will be to have a form:

\begin{displaymath}
F_{2}=(ab\vee ac\vee bc)(ce\vee cf\vee ef)(fg)(ad\vee ag\vee dg).
\end{displaymath}

Opened parenthesis and made absorptions, we obtain:

\begin{displaymath}
F=F_{1}\wedge F_{2}={\bar a}bcd{\bar e}fg\vee ab{\bar c}{\bar d}efg.
\end{displaymath}

Obviously, the expression $F$ determines two Hamiltonian cycles, each of
which contains edges whose Boolean variables have no negatives.

On the second hand, if we shall consider the theta-graph, shown on Fig.
\ref{ex} (b), we shall obtain the value $F=0$. 

\section*{Acknowledgements}

Thanks to Douglas B. West and Dan Pehoushek for useful conversations.

\end{document}